
\font\bigbf=cmbx10 scaled\magstep4
\font\seventeenbf=cmbx10 scaled\magstep3
\font\fourteenbf=cmbx10 scaled\magstep2

\font\twelvesy=cmsy10 scaled 1200
\font\twelveex=cmex10 scaled 1200
\font\twelvesc=cmcsc10 scaled 1200

\font\twelverm=cmr12
\font\twelvei=cmmi12
\font\twelvebf=cmbx12
\font\twelvesl=cmsl12
\font\twelvett=cmtt12
\font\twelveit=cmti12
\font\twelvesf=cmss12
\font\tensc=cmcsc10
\skewchar\twelvei='177
\skewchar\twelvesy='60
 \def\twelvepoint{\normalbaselineskip=12.4pt plus 0.1pt minus 0.1pt
  \abovedisplayskip 15.4pt plus 3pt minus 9pt
  \belowdisplayskip 15.4pt plus 3pt minus 9pt
  \abovedisplayshortskip 3pt plus 3pt
  \belowdisplayshortskip 10.2pt plus 3pt minus 4pt
  \smallskipamount=3.6pt plus1.2pt minus1.2pt
  \medskipamount=7.2pt plus2.4pt minus2.4pt
  \bigskipamount=14.4pt plus4.8pt minus4.8pt
  \def\rm{\fam0\twelverm}
  \def\it{\fam\itfam\twelveit}
  \def\sl{\fam\slfam\twelvesl}
  \def\bf{\fam\bffam\twelvebf}
  \def\mit{\fam 1}
  \def\cal{\fam 2}
  \def\sc{\twelvesc}
  \def\tt{\twelvett}
  \def\sf{\twelvesf}
  \textfont0=\twelverm   \scriptfont0=\tenrm   \scriptscriptfont0=\sevenrm
  \textfont1=\twelvei    \scriptfont1=\teni    \scriptscriptfont1=\seveni
  \textfont2=\twelvesy   \scriptfont2=\tensy   \scriptscriptfont2=\sevensy
  \textfont3=\twelveex   \scriptfont3=\twelveex  \scriptscriptfont3=\twelveex
  \textfont\itfam=\twelveit  \textfont\slfam=\twelvesl
  \textfont\bffam=\twelvebf  \scriptfont\bffam=\tenbf
  \scriptscriptfont\bffam=\sevenbf
  \normalbaselines\rm}

\let\medtype=\tenpoint

\def\beginlinemode{\endmode
  \begingroup\parskip=0pt \obeylines\def\\{\par}\def\endmode{\par\endgroup}}
\def\beginparmode{\endmode
  \begingroup \def\endmode{\par\endgroup}}
\let\endmode=\par
{\obeylines\gdef\
{}}
\def\singlespace{\baselineskip=\normalbaselineskip}

\def\oneandahalfspace{\baselineskip=\normalbaselineskip
  \multiply\baselineskip by 3 \divide\baselineskip by 2}
\def\doublespace{\baselineskip=\normalbaselineskip \multiply\baselineskip by 2}

\pageno=0
\newcount\firstpageno\firstpageno=2
\footline={\ifnum\pageno<\firstpageno{\hfil}\else{\hfil\twelverm\folio\hfil}\fi}
\def\toppageno{\global\footline={\hfil}\global\headline
  ={\ifnum\pageno<\firstpageno{\hfil}\else{\hfil\twelverm\folio\hfil}\fi}}
\let\rawfootnote=\footnote
\def\footnote#1#2{{\rm\singlespace\parindent=0pt\parskip=0pt
  \rawfootnote{#1}{#2\hfill\vrule height 0pt depth 6pt width 0pt}}}
\def\raggedcenter{\leftskip=3em plus 12em \rightskip=\leftskip
  \parindent=0pt \parfillskip=0pt \spaceskip=.3333em \xspaceskip=.5em
  \pretolerance=9999 \tolerance=9999
  \hyphenpenalty=9999 \exhyphenpenalty=9999 }
\def\dateline{\rightline{\ifcase\month\or
  January\or February\or March\or April\or May\or June\or
  July\or August\or September\or October\or November\or December\fi
  \space\number\year}}
\def\ldateline{\leftline{\ifcase\month\or
  January\or February\or March\or April\or May\or June\or
  July\or August\or September\or October\or November\or December\fi
  \space\number\year}}

\hsize=16.5cm
\vsize=23.0cm
\hoffset=0in
\voffset=0in
\parskip=\medskipamount
\def\\{\cr}
\twelvepoint \oneandahalfspace
\overfullrule=0pt
\newcount\timehour
\newcount\timeminute
\newcount\timehourminute
\def\daytime{\timehour=\time\divide\timehour by 60
  \timehourminute=\timehour\multiply\timehourminute by -60
  \timeminute=\time\advance\timeminute by \timehourminute
  \number\timehour:\ifnum\timeminute<10{0}\fi\number\timeminute}
\def\today{\number\day\space\ifcase\month\or Jan\or Feb\or Mar
  \or Apr\or May\or Jun\or Jul\or Aug\or Sep\or Oct\or
  Nov\or Dec\fi\space\number\year}

\def\title{\null\vskip 5pt plus 0.2fill
   \beginlinemode \doublespace \raggedcenter \bigbf}
\def\author{\vskip 3pt plus 0.2fill \beginlinemode  \doublespace
   \raggedcenter \fourteenbf}
\def\affil{\vskip 3pt plus 0.1fill
   \beginlinemode \oneandahalfspace \raggedcenter \it}

\def\abstract{\vskip 3pt plus 0.3fill {\raggedcenter{\rm ABSTRACT}}
   \beginparmode \narrower \oneandahalfspace }
\def\body{\beginparmode}
\def\endtopmatter{\endpage\body}

\def\head#1{ \vfill\eject\null\vskip 0.25in  {\immediate\write16{#1}
   \raggedcenter {\seventeenbf #1} \par \bigskip}  
   \nobreak\vskip 0in\nobreak}
\def\subhead#1{\vskip 0.25in
  {\leftline {\bf #1} \par}
   \nobreak\vskip 0in\nobreak}
\def\subsubhead#1{ \vskip 0.25in  {\leftline {\bf #1} \par}
   \nobreak\vskip 0in\nobreak}

\newcount\figno\figno=0
\newcount\chapno\chapno=0
\newcount\subchapno\subchapno=0
\newcount\subsubchapno\subsubchapno=0

\outer\def\chap#1{\global\advance\chapno by 1
\global\figno=0
\global\subchapno=0 \global\subsubchapno=0
 \head{\the \chapno. #1} \taghead{\the\chapno.} }

\outer\def\subchap#1{\global\advance\subchapno by 1
 \global\subsubchapno=0
 \subhead{\the\chapno.\the\subchapno. #1}  }

\outer\def\exer{\global\advance\subchapno by 1
 \global\subsubchapno=0
 \subhead{Exercice \the\chapno.\the\subchapno. }  }

\outer\def\subsubchap#1{\global\advance\subsubchapno by 1
 \subsubhead{\the\chapno.\the\subchapno.\the\subsubchapno. #1} }

\outer\def\Fig #1 #2 #3 #4 #5{\global\advance\figno by 1
\midinsert \vglue#1cm \hskip#2cm
\special{picture #3 scaled #4} \medskip
\medtype \singlespace \parshape=2 2.25cm 11cm 3.75cm 9.5cm
{ {\bf Figure \the\chapno.\the\figno.}  \rm #5.}  \endinsert}

\outer\def\Fign #1 #2 #3 #4 {\global\advance\figno by 1
\midinsert  \vglue#1cm \hskip#2cm
\special{picture #3 scaled #4} \medskip
{\centerline{\tenbf Figure \the\chapno.\the\figno. }}
\endinsert}

\def\references{\head{References}
   \beginparmode\singlespace    
   \frenchspacing \parindent=30pt
      \parskip=\smallskipamount 
 \everypar={\hangindent=40pt  \hangafter=1}}
\def\referencesweird{
    \beginparmode\singlespace  \frenchspacing \parindent=30pt
      \parskip=\smallskipamount
 \everypar={\hangindent=40pt  \hangafter=1}}

\def\refstyleprnp{                
  \gdef\refto##1{$\rm [##1]$}
  \gdef\refis##1{\item{[##1]\ }}
  \gdef\journal##1,##2,##3,##4&{{\sl ##1 }{\bf ##2 }(##4) ##3}}

\refstyleprnp  

\def\cmp{\journal Comm. Math. Phys.}

\def\ijmp{\journal Int. J. Mod. Phys.}

\def\np{\journal Nucl. Phys.}
\def\pl{\journal Phys. Lett.}

\def\endreferences{\body}

\def\ref#1{Ref.~#1}               
\def\Ref#1{Ref.~[#1]}
\def\[#1]{[\cite{#1}]}
\def\cite#1{{#1}}


\def\(#1){(\call{#1})}
\def\call#1{{#1}}

\def\taghead#1{}

\def\beneathrel#1\under#2{\mathrel{\mathop{#2}\limits_{#1}}}
\def\frac#1#2{{#1 \over #2}}
\def\13{{1\over3}}
\def\14{{1\over4}}
\def\12{{1\over2}}

\def\ie{{\it i.e. }}

\def\sla{\raise.15ex\hbox{$/$}\kern -.57em}
\def\leaderfill{\leaders\hbox to 1em{\hss.\hss}\hfill}
\def\twiddle{\lower.9ex\rlap{$\kern -.1em\scriptstyle\sim$}}
\def\bigtwiddle{\lower1.ex\rlap{$\sim$}}
\def\gtwid{\mathrel{\raise.3ex\hbox{$>$\kern -.75em\lower1ex\hbox{$\sim$}}}}
\def\ltwid{\mathrel{\raise.3ex\hbox{$<$\kern -.75em\lower1ex\hbox{$\sim$}}}}
\def\tdot#1{\mathord{\mathop{#1}\limits^{\kern2pt\ldots}}}
\def\pmb#1{\setbox0=\hbox{#1}   
  \kern -.025em\copy0\kern -\wd0
  \kern  .05em\copy0\kern -\wd0
  \kern -.025em\raise.0433em\box0 }
\def\square{\kern1pt\vbox{\hrule height 0.6pt\hbox{\vrule width 0.6pt\hskip 3pt
   \vbox{\vskip 6pt}\hskip 3pt\vrule width 0.6pt}\hrule height 0.6pt}\kern1pt}

\def\ket#1{\left| #1\right\rangle}

\def\rchi{\raise2pt\hbox{$\chi$}}

\def\Re{{\cal R \mskip-4mu \lower.1ex \hbox{\it e}}}
\def\Im{{\cal I \mskip-5mu \lower.1ex \hbox{\it m}}}
\def\tr{{\rm tr\,}}

\def\rlh{\scriptstyle{\rightharpoonup\hskip-8pt{\leftharpoondown}}}
\def\harpar{\partial\hskip-8pt\raise9pt\hbox{$\rlh$}}

\def\Zint{\ {{\bf Z} \kern -.40em {\bf Z}}}
\def\real{{\vrule height 1.5ex width 0.05em depth 0ex \kern
-0.06em {\bf R}}}
\def\natural{\ {\vrule height 1.5ex width 0.05em depth 0ex \kern
-0.06em {\bf N}}}
\def\complex{\ {\vrule height 1.4ex width 0.05em depth 0ex \kern
-0.29em {\bf C}}}
\def\rational{\ {\vrule height 1.4ex width 0.05em depth 0ex \kern
-0.29em {\bf Q}}}

\def\diagram{\def\normalbaselines{\baselineskip20pt \lineskip3pt
\lineskiplimit3pt}}

\def\endpage {\vfill\eject}
\def\endpaper{\endmode\vfill\supereject}
\def\endit {\endpaper\end}
\catcode`@=11
\newcount\tagnumber\tagnumber=0
\immediate\newwrite\eqnfile
\newif\if@qnfile\@qnfilefalse
\def\write@qn#1{}
\def\writenew@qn#1{}
\def\w@rnqwrite#1{\write@qn{#1}\message{#1}}
\def\@rrwrite#1{\write@qn{#1}\errmessage{#1}}

\def\taghead#1{\gdef\t@ghead{#1}\global\tagnumber=0}
\def\t@ghead{}

\expandafter\def\csname @qnnum -3\endcsname
  {{\t@ghead\advance\tagnumber by -3\relax\number\tagnumber}}
\expandafter\def\csname @qnnum -2\endcsname
  {{\t@ghead\advance\tagnumber by -2\relax\number\tagnumber}}
\expandafter\def\csname @qnnum -1\endcsname
  {{\t@ghead\advance\tagnumber by -1\relax\number\tagnumber}}
\expandafter\def\csname @qnnum0\endcsname
  {\t@ghead\number\tagnumber}
\expandafter\def\csname @qnnum+1\endcsname
  {{\t@ghead\advance\tagnumber by 1\relax\number\tagnumber}}
\expandafter\def\csname @qnnum+2\endcsname
  {{\t@ghead\advance\tagnumber by 2\relax\number\tagnumber}}
\expandafter\def\csname @qnnum+3\endcsname
  {{\t@ghead\advance\tagnumber by 3\relax\number\tagnumber}}

\def\equationfile{%
  \@qnfiletrue\immediate\openout\eqnfile=\jobname.eqn%
  \def\write@qn##1{\if@qnfile\immediate\write\eqnfile{##1}\fi}
  \def\writenew@qn##1{\if@qnfile\immediate\write\eqnfile
     {\noexpand\tag{##1} = (\t@ghead\number\tagnumber)}\fi}
     }
\def\callall#1{\xdef#1##1{#1{\noexpand\call{##1}}}}
\def\call#1{\each@rg\callr@nge{#1}}

\def\each@rg#1#2{{\let\thecsname=#1\expandafter\first@rg#2,\end,}}
\def\first@rg#1,{\thecsname{#1}\apply@rg}
\def\apply@rg#1,{\ifx\end#1\let\next=\relax%
\else,\thecsname{#1}\let\next=\apply@rg\fi\next}

\def\callr@nge#1{\calldor@nge#1-\end -}
\def\callr@ngeat#1\end -{#1}
\def\calldor@nge#1-#2-{\ifx\end#2\@qneatspace#1 %
  \else\calll@@p{#1}{#2}\callr@ngeat\fi}
\def\calll@@p#1#2{\ifnum#1>#2{\@rrwrite{Equation range #1-#2\space is bad.}
\errhelp{If you call a series of equations by the notation M-N, then M and
N must be integers, and N must be greater than or equal to M.}}\else%
 {\count0=#1\count1=#2\advance\count1
by1\relax\expandafter\@qncall\the\count0,%
  \loop\advance\count0 by1\relax%
    \ifnum\count0<\count1,\expandafter\@qncall\the\count0,%
  \repeat}\fi}

\def\@qneatspace#1#2 {\@qncall#1#2,}
\def\@qncall#1,{\ifunc@lled{#1}{\def\next{#1}\ifx\next\empty\else
  \w@rnqwrite{Equation \noexpand\(>>#1<<) not defined yet.}
  >>#1<<\fi}\else\csname @qnnum#1\endcsname\fi}

\let\eqnono=\eqno
\def\eqno(#1){\tag#1}
\def\tag#1$${\eqnono(\displayt@g#1 )$$}

\def\aligntag#1\endaligntag
  $${\gdef\tag##1\\{&(##1 )\cr}\eqalignno{#1\\}$$
  \gdef\tag##1$${\eqnono(\displayt@g##1 )$$}}

\def\eqalignno#1{\displ@y \tabskip\centering
  \halign to\displaywidth{\hfil$\displaystyle{##}$\tabskip\z@skip
    &$\displaystyle{{}##}$\hfil\tabskip\centering
    &\llap{$\displayt@gpar##$}\tabskip\z@skip\crcr
    #1\crcr}}

\def\displayt@gpar(#1){(\displayt@g#1 )}
\def\displayt@g#1 {\rm\ifunc@lled{#1}\global\advance\tagnumber by1
        {\def\next{#1}\ifx\next\empty\else\expandafter
        \xdef\csname @qnnum#1\endcsname{\t@ghead\number\tagnumber}\fi}%
  \writenew@qn{#1}\t@ghead\number\tagnumber\else
        {\edef\next{\t@ghead\number\tagnumber}%
        \expandafter\ifx\csname @qnnum#1\endcsname\next\else
        \w@rnqwrite{Equation \noexpand\tag{#1} is a duplicate number.}\fi}%
  \csname @qnnum#1\endcsname\fi}

\def\ifunc@lled#1{\expandafter\ifx\csname @qnnum#1\endcsname\relax}
\let\@qnend=\end\gdef\end{\if@qnfile
\immediate\write16{Equation labels written on []\jobname.eqn.}\fi\@qnend}
\newcount\r@fcount \r@fcount=0
\newcount\r@fcurr
\immediate\newwrite\reffile
\newif\ifr@ffile\r@ffilefalse
\def\w@rnwrite#1{\ifr@ffile\immediate\write\reffile{#1}\fi\message{#1}}
\def\writer@f#1>>{}
\def\referencefile{
  \r@ffiletrue\immediate\openout\reffile=\jobname.ref%
  \def\writer@f##1>>{\ifr@ffile\immediate\write\reffile%
    {\noexpand\refis{##1} =\csname r@fnum##1\endcsname = %
     \expandafter\expandafter\expandafter\strip@t\expandafter%
     \meaning\csname r@ftext\csname r@fnum##1\endcsname\endcsname}\fi}%
  \def\strip@t##1>>{}}

\def\citeall#1{\xdef#1##1{#1{\noexpand\cite{##1}}}}
\def\cite#1{\each@rg\citer@nge{#1}}     

\def\each@rg#1#2{{\let\thecsname=#1\expandafter\first@rg#2,\end,}}
\def\first@rg#1,{\thecsname{#1}\apply@rg}         
\def\apply@rg#1,{\ifx\end#1\let\next=\relax       
\else,\thecsname{#1}\let\next=\apply@rg\fi\next}  

\def\citer@nge#1{\citedor@nge#1-\end -}     
\def\citer@ngeat#1\end -{#1}
\def\citedor@nge#1-#2-{\ifx\end#2\r@featspace#1 
  \else\citel@@p{#1}{#2}\citer@ngeat\fi}     
\def\citel@@p#1#2{\ifnum#1>#2{\errmessage{Reference range #1-#2\space is bad.}
    \errhelp{If you cite a series of references by the notation M-N, then M and
    N must be integers, and N must be greater than or equal to M.}}\else%
 {\count0=#1\count1=#2\advance\count1
by1\relax\expandafter\r@fcite\the\count0,%
  \loop\advance\count0 by1\relax
    \ifnum\count0<\count1,\expandafter\r@fcite\the\count0,%
  \repeat}\fi}

\def\r@featspace#1#2 {\r@fcite#1#2,}     
\def\r@fcite#1,{\ifuncit@d{#1}          
    \expandafter\gdef\csname r@ftext\number\r@fcount\endcsname%
    {\message{Reference #1 to be supplied.}\writer@f#1>>#1 to be supplied.\par
     }\fi%
  \csname r@fnum#1\endcsname}

\def\ifuncit@d#1{\expandafter\ifx\csname r@fnum#1\endcsname\relax%
\global\advance\r@fcount by1%
\expandafter\xdef\csname r@fnum#1\endcsname{\number\r@fcount}}

\let\r@fis=\refis                         
\def\refis#1#2#3\par{\ifuncit@d{#1}
    \w@rnwrite{Reference #1=\number\r@fcount\space  not yet cited }\fi%
  \expandafter\gdef\csname r@ftext\csname r@fnum#1\endcsname\endcsname%
  {\writer@f#1>>#2#3\par}}

\def\r@ferr{\endreferences\errmessage{There should have been
\noexpand\endreferences before now; it has been inserted here.}}

\let\r@ferences=\references
\def\references{\r@ferences\def\endmode{\r@ferr\par\endgroup}}
\let\r@ferencesweird=\referencesweird
\def\refweird#1{\r@fcount=#1\r@ferencesweird\def\endmode{\r@ferr\par\endgroup}}

\let\endr@ferences=\endreferences
\def\endreferences{\r@fcurr=0
  {\loop\ifnum\r@fcurr<\r@fcount
    \advance\r@fcurr by 1\relax\expandafter\r@fis\expandafter{\number\r@fcurr}%
    \csname r@ftext\number\r@fcurr\endcsname%
  \repeat}\gdef\r@ferr{}\endr@ferences}
\def\endrefweird#1{\r@fcurr=#1%
  {\loop\ifnum\r@fcurr<\r@fcount%
    \advance\r@fcurr by 1\relax\expandafter\r@fis\expandafter{\number\r@fcurr}%
    \csname r@ftext\number\r@fcurr\endcsname%
  \repeat}\gdef\r@ferr{}\endr@ferences}
\let\r@fend=\endpaper\gdef\endpaper{\ifr@ffile
\immediate\write16{References written on []\jobname.ref.}\fi\r@fend}
\catcode`@=12
\citeall\refto \citeall\ref \citeall\Ref
\catcode`@=11
\immediate\newwrite\tocfile
 \newif\ift@cs\t@csfalse
\def\tocs{\t@cstrue\immediate\openout\tocfile=\jobname.toc%
\newlinechar=`^^J %
\write\tocfile{
  \string\pageno=-1\string\firstpageno=-1000\string\singlespace %
  \string\null\string\vfill\string\centerline{TABLE OF CONTENTS}^^J %
  \string\vskip 0.5in\string\rightline{\string\underbar{Page}}\smallskip}}

\def\tocitem#1{
  \t@cskip{#1}\bigskip}
\def\tocitemitem#1{
  \t@cskip{\quad#1}\medskip}
\def\tocitemitemitem#1{
  \t@cskip{\qquad#1}\smallskip}
\def\tocitemall#1{
  \xdef#1##1{#1{##1}\noexpand\tocitem{##1}}}
\def\tocitemitemall#1{
  \xdef#1##1{#1{##1}\noexpand\tocitemitem{##1}}}
\def\tocitemitemitemall#1{
  \xdef#1##1{#1{##1}\noexpand\tocitemitemitem{##1}}}

\def\t@cskip#1#2{
  \write\tocfile{\string#2\string\line{^^J
  #1\string\leaderfill\space\number\folio}}}
\def\t@cproduce{
  \write\tocfile{\string\vfill\string\vfill\string\supereject\string\end}
  \closeout\tocfile
  \immediate\write16{Table of Contents written on []\jobname.toc.}}
\let\t@cend=\endpaper\def\endpaper{\ift@cs\t@cproduce\fi\t@cend}
\catcode`@=12
\tocitemall\head          
\tocitemitemall\subhead
\tocitemitemitemall\subsubhead

$\quad$\vskip7cm\vfill

\title{New$\quad$Solutions$\quad$to$\quad$the
Yang--Baxter$\quad$Equation$\quad $from
Two--Dimensional$\quad$Representations
of$\quad$U{\lower7pt\hbox{q}}(sl(2))$\quad$at$\quad$Roots$\quad$of$\quad$Unit}

\vskip.5cm

\author{M.~Ruiz--Altaba$^*$}

\affil{D\'ept. Physique Th\'eorique,  Universit\'e de %
Gen\`eve,  CH--1211  Gen\`eve 4
\footnote{}{$^*$Supported in part by the  Fonds National %
Suisse pour la Recherche Scientifique.}}

\affil{ \rm{UGVA--DPT~1991/08--741} }
\vskip1cm

\beginparmode\narrower{{\bf Abstract}: We  present
 particularly simple  new solutions to the Yang--Baxter
equation arising from two--dimensional cyclic
representations of quantum $SU(2)$. They are readily
interpreted as scattering matrices of relativistic
objects, and the quantum group becomes a dynamical
symmetry.}

\endtopmatter\pageno=1

\subhead{1. Introduction: $U_\epsilon(s\ell(2))$ with %
$\epsilon^4=1$} \taghead{1.}

Quantum groups at roots of unit enjoy a beautiful
representation theory \refto{5,4,6,Ros,12,9} which has
been applied succesfully to the understanding of the
chiral Potts model \refto{1,2,4,3,8,88}.
In this letter, we apply the general formalism of cyclic
representations  of $U_\epsilon(s\ell(2))$
to the somewhat degenerate case of $\epsilon^4=1$, \ie
$q^2=1$. It complements the work for $U_q(s\ell(2))$ in
\refto{6} of the $p=3$ case  and in \refto{12} of the limit
$p\to\infty$, as well as the case $p=2$ of $U_q(s\ell(3))$
in \refto{33}.

 When $q^p=1$, the center of
$U_\epsilon(s\ell(2))$ contains not only the standard
quadratic Casimir $$C=K^{-1}FE +{1\over q-1} \left(
q^{-1}K+K^{-1} \right) \tag $$ but also the elements $F^p$,
$E^p$ and $K^p$.
 We work in the contour basis \refto{11}  with
co-multiplication  $$\eqalign{ &\Delta E = E \otimes {\bf
1} + K \otimes E \cr
 &\Delta F = F \otimes {\bf 1} + K \otimes F \cr
&\Delta K = K\otimes K \cr} \eqno(4)$$
The
generators $E$, $F$ and $K$ of $U_\epsilon(s\ell(2))$
satisfy  the standard  relations : $$\eqalign{ &EF-qFE=
{1-K^2 }\cr & EK-qKE=0 \cr & FK-q^{-1}KF=0 \cr}\eqno(5)$$
Let us denote by $\xi={(x,y,z)}  $  the eigenvalues of
$(E^p,F^p,K^p)$, and introduce the notation $$\eqalign{
\mu= {x\over 1-z} \cr \nu= {y\over 1-z} \cr}\tag$$

Specializing to the case of interest, namely $p=2$
($q=-1$), and letting
$z=\lambda^2$, the Casimir eigenvalue is then  $$c=
 {\left( 2- \sqrt{ 1-4\mu\nu} \right) \over2} \left(
\lambda-\lambda^{-1} \right)\tag$$ The spectrum of
$U_\epsilon(s\ell(2))$ with $\epsilon= e^{i\pi/2}$ consists
of a three--dimensional continuum of two--dimensional
representations,  labelled by $\xi$, with a singular
(orbifold)  point at $z=1$, which corresponds to
the only regular representation in this theory, namely the
identity.

The irreducible representations of the theory
consist thus the identity operator and the manifold of
doublet cyclic representations. The latter constitute an
intrinsically quantum generalization of the customary and
useful doublet irrep of $SU(2)$, and the purpose of this
letter is to analyze some of their physical properties
from the purely algebraic point of view.

Let $e_r(\xi)$ ($r=0,1$) be the basis for the cyclic
representation $\pi_\xi=\pi_{(x,y,z=\lambda^2)}$,  defined
as follows ($z=\lambda^2\not=1$):  $$\diagram\matrix{
Ke_0=\lambda e_0  \qquad &Fe_0=\sqrt{\nu}\left( 1-\lambda
\right) e_1 \qquad & Ee_0 = {1-\sqrt{ 1-4\mu\nu} \over
2\sqrt{\nu}} (1-\lambda) e_1 \cr Ke_1=-\lambda e_1
  \qquad &Fe_1=\sqrt{\nu}\left( 1+\lambda \right) e_0
\qquad& Ee_1= {1+\sqrt{ 1-4\mu\nu} \over 2\sqrt{\nu}}
(1+\lambda) e_0\cr}\eqno(3)$$ In our choice of basis, we
have implicitly assumed that $\nu\not=0$. The special class
of representations with $\mu=0$ are called semi-cyclic: for
them, $e_0$ is a highest weight ($Ee_0=0$) and yet $e_1$ is
not a lowest weight ($Fe_1\not=0$).

 The seasoned reader is certainly struck by the appearence
of anti-commutators in the relations among the generators.
Quite simply, when $q^2=1$, $q$-commutators become
anti-commutators. For future reference, let us introduce a
``fermionic'' basis for $U_\epsilon(s\ell(2))$ when
$\epsilon^4=1$: $$\eqalign{ & b= {1\over 1+K} E \cr &
b^\dagger = {1\over 1+K} F \cr}\tag$$  The quantum algebra is then generated by
$K$,
$b$ and $b^\dagger$, with the following  anticommutation
relations: $$\eqalign{ & \left\{ b,K\right\} = \left\{
b^\dagger,K\right\} = 0 \cr
 & \left\{ b , b^\dagger  \right\} =1 \cr
 & \left\{ b , b  \right\} =2\mu \cr
 & \left\{ b^\dagger , b^\dagger  \right\} =2\nu \cr }\tag$$
($\mu$ and $\nu$ are
$c$--numbers or, equivalently, operators proportional to
the identity). We may rescale the quadratic Casimir by
$(K^2-1)$ to get the quantum relative of the number
operator:  $$Q=K  \left( b^\dagger b
-\12 \right) \tag$$ which commutes with $K$, $b$ and
$b^\dagger$.
 A cyclic irreducible representation is characterized in
this language by $e_0=\ket-$ and $e_1=\ket+$ with
$$\eqalign{ &K\ket\pm = \pm \lambda \ket\pm \cr
 & b \ket \pm = {1\mp \sqrt{1-4\mu\nu} \over 2\sqrt{\nu}}
\ket\mp \cr
 & b^\dagger \ket \pm = \sqrt{\nu} \ket\mp \cr}$$
 Again, note that in the semi-cyclic case, $\mu=0$, we may
think of $\ket+$ as akin to
 the ground state (it is annihilated by $b$).

\subhead{2. Intertwiners and Yang--Baxter equation: general
solution} \taghead{2.}

The intertwiner $R(\xi_1,\xi_2)$ between two cyclic
representations effects a braiding and can be thought of
as a $2\to2$ scattering matrix:  $$e_{r_1}(\xi_1) \otimes
e_{r_2}(\xi_2) = R_{r_1 r_2}^{r'_1r'_2} (\xi_1, \xi_2)
e_{r'_1}(\xi_2) \otimes e_{r'_2}(\xi_1) \eqno(6)$$
Quasi--triangularity requires $[R, \Delta]=0$, \ie  $$
R_{r'_1 r'_2}^{r''_1r''_2} (\xi_1, \xi_2)
\Delta_{\xi_1,\xi_2} (g) _{r_1 r_2}^{r'_1r'_2}  =
\Delta_{\xi_2,\xi_1} (g)_{r'_1 r'_2}^{r''_1r''_2} R_{r_1
r_2}^{r'_1r'_2} (\xi_1, \xi_2) \eqno(7)$$ for any $g\in
U_q(s\ell(2))$. Specializing to $g=E^2$ and $g=F^2$, we see
that if the intertwiner $R(\xi_1,\xi_2)$ is to exist,
then   $\xi_1$ and $\xi_2$ are constrained to lie on the
same spectral variety~\refto{4}:  $$\eqalign{& {x_1
\over 1-\lambda^2_1 }={x_2 \over 1-\lambda^2_2 }= \mu \cr&
{y_1 \over 1-\lambda^2_1 }={y_2 \over 1-\lambda^2_2 }= \nu
\cr} \eqno()$$ with arbitrary $\mu, \nu \in {\bf C}$ and
$\lambda_i^2\not=1$.

  The main result we discuss in this letter is that, for
$\xi_1$, $\xi_2$, $\xi_3$ on the same spectral variety
  there exists  an $R$--matrix $R(\xi_i,\xi_j)$
satisfying~\(7) and the Yang--Baxter equation
$$\sum_{s_1,s_2,s_3} R_{s_1 s_3}^{r'_1 r'_2} (\xi_2,
\xi_3)  R_{s_2 r_3}^{s_3 r'_3} (\xi_1, \xi_3) R_{r_1
r_2}^{s_1s_2} (\xi_1, \xi_2) =  \sum_{s_1, s_2, s_3} R_{s_1
s_3}^{r'_2 r'_3} (\xi_1, \xi_2)  R_{r_1 s_2}^{r'_1 s_1}
(\xi_1, \xi_3)  R_{r_2 r_3}^{s_2 s_3} (\xi_2, \xi_3)
\eqno(9)$$

 The explicit form of the $R$--matrix intertwiner
satisfying the Yang--Baxter equation is the following (we
show only the non--zero entries): $$\eqalign{
& R_{00}^{00}(\xi_1, \xi_2)=1 \cr
 & R_{01}^{01}(\xi_1,\xi_2)=\Omega_1
{\left(1-\lambda_1 \right) \left( 1+ \lambda_2\right) \over
 \Omega_2 -\lambda_1\lambda_2 \Omega_1} \cr
& R_{01}^{10}(\xi_1, \xi_2)={ \lambda_1\Omega_2 - \lambda_2 \Omega_1 \over
\Omega_2 - \lambda_1 \lambda_2 \Omega_1}\cr
& R_{10}^{01}(\xi_1, \xi_2)= { \lambda_2\Omega_2 - \lambda_1 \Omega_1 \over
\Omega_2 - \lambda_1 \lambda_2 \Omega_1} \cr
& R_{10}^{10}(\xi_1, \xi_2)= \Omega_2 { \left(1+\lambda_1 \right) \left( 1-
\lambda_2\right) \over \Omega_2 -\lambda_1\lambda_2 \Omega_1} \cr
& R_{11}^{11}(\xi_1, \xi_2)= { \Omega_1 - \lambda_1 \lambda_2 \Omega_2 \over
\Omega_2 - \lambda_1 \lambda_2 \Omega_1}\cr}\eqno(rmat)$$
Here, $\Omega_i=\Omega(\xi_i)$ are the values of an arbitrary function of the
labels of the cyclic representation. Note that what really appears is only the
ratio $\Omega_1/\Omega_2= \Omega(\lambda_1)/\Omega(\lambda_2)$.

The  $R$--matrix \(rmat) with arbitary $\Omega(\xi)$
enjoys two remarkable properties. First, it is normalized:
$$R_{r_1 r_2}^{r'_1 r'_2} (\xi,\xi) = \delta _{r_1}^{r'_1}
\delta_{r_2}^{r'_2} \eqno(11)$$ Second, it is unitary:
$$\sum_{r'_1 r'_2} R_{r_1 r_2}^{r'_1 r'_2} (\xi_1 , \xi_2) R_{r'_1
r'_2}^{r''_1 r''_2} (\xi_2, \xi_1) = \delta _{r_1}^{r''_1} \delta
_{r_2}^{r''_2} \eqno(12)$$

 Let us stress some important differences between the above
solution for $q^2=1$ and the generic ($q^p=1$, $p\ge3$)
semi-cyclic ($y=0$) situation \refto{6,12}. Firstly,
this is the only case in which an arbitrary function
$\Omega$ appears, \ie there is a whole family of
spectral--dependent $R$--matrices. Secondly, the
$R$--matrix \(rmat) does not involve the parameters $\mu=
x_i/(1-z_i)$ nor $\nu= y_i/(1-z_i)$ explicitly, although of
course $\xi_1$ and $\xi_2$ must lie on the same spectral
variety, \ie share common values for $\mu$ and $\nu$, for
$R(\xi_1,\xi_2)$ to exist at all. (A dependence of $R$ on
$\mu$ and $\nu$ could be introduced through $\Omega$.)
Thirdly, the $R$--matrix \(rmat) conserves the ``quantum
isospin'' exactly $$R_{r_1, r_2}^{r'_1, r'_2} (\xi_1,
\xi_2) \not=0 \Longrightarrow r_1+r_2 = r'_1+r'_2 \tag $$
instead of just modulo 2. In particular, $R_{11}^{00}$ and
$R_{00}^{11}$ are always zero, so if we interpret the
$R$--matrix entries as Boltzmann weights, we are dealing
with a six--vertex model. These three observations are not
unrelated, as will become clearer below.

 The arbitrary function $\Omega$ is quite akin to an
affinization parameter. This interpretation is reinforced
by the fact that only the ratio $\Omega_1/\Omega_2$ appears
in the $R$--matrix, which can be decomposed as follows
$$R(\lambda_1, \lambda_2; \Omega_1/\Omega_2) = { \left(
\lambda_1 \lambda_2 {\Omega_1 \over \Omega_2} \right)
^{-1/2} R(\lambda_1,\lambda_2; 0) -  \left( \lambda_1
\lambda_2 {\Omega_1 \over \Omega_2} \right) ^{1/2}
R(\lambda_1,\lambda_2; \infty) \over  \left( \lambda_1
\lambda_2 {\Omega_1 \over \Omega_2} \right) ^{-1/2}  -
\left( \lambda_1 \lambda_2 {\Omega_1 \over \Omega_2}
\right) ^{1/2}} \tag $$ or, less symmetrically but more
succinctly, as  $$R(\lambda_1, \lambda_2;
\Omega_1/\Omega_2) = {1\over  1  -  \lambda_1 \lambda_2
{\Omega_1 \over \Omega_2} }
 \left[  R(\lambda_1,\lambda_2; 0) -  \lambda_1 \lambda_2
{\Omega_1 \over \Omega_2} R(\lambda_1,\lambda_2; \infty) \right] \tag $$
 We may thus view $R(\lambda_1,\lambda_2;0)$  as the
 basic $R$--matrix from which the above family is
built, because $R(\lambda_1,\lambda_2;\infty)=
R(\lambda_1,\lambda_2;0)^{-1}$. Remarkably, the
$R$--matrix $R(\lambda_1,\lambda_2;0)$ can be obtained with
the help of contour techniques in the
semi-cyclic case ($\mu=0$) supplemented with the rule that
$F^2=0$. It is thus apparent that
$R(\lambda_1,\lambda_2;0)$ is a slight generalization of
the usual $R^{\12\12}$,   the $R$--matrix for the
regular doublet representation of  quantum $SU(2)$ with
$q^p=1$ ($p\ge3$). In fact, $R(\lambda_1,\lambda_2;0)$ has
been  considered in \refto{Ros}. Let us conclude this
detour by  noting that we may not give a contour
representation of  the general
$R(\lambda_1,\lambda_2;\Omega)$: since it  is a linear
 combination of $R(\lambda_1,\lambda_2;0)$ and
$R(\lambda_1,\lambda_2;0)^{-1}$, it consists of a piece due
to braiding by $\pi$ and another by $-\pi$.

\subhead{3. Clebsch--Gordan coefficients and crossing
symmetry} \taghead{3.}
We may consider the tensor product of two cyclic
representations $\pi_{\xi_1}$, $\pi_{\xi_2}$ on the same
spectral variety, parametrized by $(\mu,\nu)$. The result is
a direct sum of two  cyclic representations
$\pi_{\xi_\pm}$ again on the same variety, with
$$\lambda_\pm=\pm \lambda_1 \lambda_2$$ and  $$x_\pm=
\mu(1-\lambda_\pm^2)=x_1+x_2 \lambda_1^2=x_2+x_1
\lambda_2^2$$ and similarly for $y$. It is easy to check
that   $$\eqalign{& e_0(\lambda_1\lambda_2) = e_0
(\lambda_1) \otimes e_0(\lambda_2) \cr &
e_1(\lambda_1\lambda_2) =  {-\lambda_1 \left(
1+\lambda_2 \right) \over 1-\lambda_1 \lambda_2} e_1
(\lambda_1) \otimes e_0(\lambda_2) +{1+\lambda_1
\over 1-\lambda_1 \lambda_2}
 e_0 (\lambda_1) \otimes e_1(\lambda_2) \cr
& e_0(-\lambda_1\lambda_2) = {1-\lambda_1
\over 1-\lambda_1 \lambda_2} e_1 (\lambda_1) \otimes
e_0(\lambda_2) + {\lambda_1 \left(
1-\lambda_2 \right) \over 1-\lambda_1 \lambda_2}
e_0 (\lambda_1) \otimes e_1(\lambda_2) \cr &
e_1(-\lambda_1\lambda_2) = e_1(\lambda_1) \otimes
e_1(\lambda_2) \cr}\tag $$

Thus,  the non--zero  quantum Clebsch--Gordan coefficients
are $$\eqalign{
 & K _{\lambda_1\ \lambda_2\ 0}^{0\ 0\ \lambda_1\lambda_2}
= 1 \cr
& K _{\lambda_1\ \lambda_2\ 1}^{0\ 1\ \lambda_1\lambda_2} =
{ 1+ \lambda_1 \over 1-\lambda_1 \lambda_2 }  \cr
& K _{\lambda_1\ \lambda_2\ 1}^{1\ 0\ \lambda_1\lambda_2} =
{-\lambda_1 \left( 1+ \lambda_2 \right) \over 1-\lambda_1
\lambda_2 } \cr
& K _{\lambda_1\ \lambda_2\ 0}^{0\ 1\ -\lambda_1\lambda_2}
= {\lambda_1 \left( 1- \lambda_2 \right) \over 1-\lambda_1
\lambda_2 } \cr
& K _{\lambda_1\ \lambda_2\ 0}^{1\ 0\
-\lambda_1\lambda_2} = { 1- \lambda_1 \over 1-\lambda_1
\lambda_2 }  \cr
& K _{\lambda_1\ \lambda_2\ 1}^{1\ 1\
-\lambda_1\lambda_2} = 1 \cr}\tag $$
 For completeness, we note also the
non--zero inverse quantum Clebsch--Gordan  coefficients:
$$\eqalign{
& \tilde{K} ^{\lambda_1\ \lambda_2\ 0}_{0\ 0
\ \lambda_1\lambda_2} = 1 \cr
& \tilde{K} ^{\lambda_1\ \lambda_2\ 1}_{0\ 1
\ \lambda_1\lambda_2} = {1-\lambda_1 \over
1+\lambda_1\lambda_2 }  \cr
& \tilde{K} ^{\lambda_1\ \lambda_2\ 1}_{1\ 0
\ \lambda_1\lambda_2} = {-\lambda_1 \left( 1- \lambda_2
\right)    \over 1+\lambda_1\lambda_2 }\cr
& \tilde{K} ^{\lambda_1\ \lambda_2\ 0}_{0\ 1\ -
\lambda_1\lambda_2} =  {\lambda_1 \left( 1+ \lambda_2
\right)    \over 1+\lambda_1\lambda_2 }\cr
& \tilde{K} ^{\lambda_1\ \lambda_2\ 0}_{1\ 0
\   -\lambda_1\lambda_2} = {1+\lambda_1 \over
1+\lambda_1\lambda_2 }  \cr
& \tilde{K} ^{\lambda_1\
\lambda_2\ 1} _{1\ 1\ -\lambda_1\lambda_2}
= 1 \cr}\tag $$

Imagine now decomposing  $e_i(\pm\lambda_1\lambda_2)$ with
$K_{\lambda_1\ \lambda_2 \  i}^{j_1\  j_2 \
\pm\lambda_1\lambda_2}$, and then   braiding the result
with $R(\lambda_1,\lambda_2;\Omega_1/\Omega_2 )$. We may
compare the result of these two actions, $R K^{12}$, with
the single immediate decomposition into $e_{j_1}(\lambda_2)
\otimes e_{j_2} (\lambda_1)$ with the help of $K_{\lambda_2
\  \lambda_1\ i}^{j_1   \  j_2 \  \pm\lambda_1\lambda_2}$,
\ie $ K^{21}$. The two operations are related by a factor,
depending only on the label of the composed representation,
$\pm\lambda_1\lambda_2$,  but not on any of the quantum
group indices $i,j\in\Zint_2$:  $$R(\lambda_1,\lambda_2;
\Omega_1/\Omega_2) _{i_1\ i_2}^{j_1\ j_2}
 K_{\lambda_1\ \lambda_2\ k}^{i_1\ i_2\ \pm
\lambda_1\lambda_2}  =  \Phi_\pm(\lambda_1,\lambda_2)
K_{\lambda_2\ \lambda_1\ k}^{j_1\ j_2\ \pm
\lambda_1\lambda_2}  \tag $$ and
$$\Phi_+(\lambda_1,\lambda_2)=1\quad, \qquad
\Phi_-(\lambda_1,\lambda_2) =
R_{11}^{11}(\lambda_1,\lambda_2;\Omega_1/\Omega_2) \tag $$

 It is interesting to note that a single unique particular
choice of the arbitrary function $\Omega$ allows us to set
$\Phi_-=1$, namely $$\Omega_1/\Omega_2 = 1
\tag omeg$$ Then, the $R$--matrix
$R(\lambda_1,\lambda_2;1)$ is given merely by the mismatch
in the decompositions of $e(\lambda_1) \otimes
e(\lambda_2)$ and   $e(\lambda_2) \otimes e(\lambda_1)$
into cyclic irreps, without any extra phase factors:
$$R_{r_1r_2}^{r'_1r'_2} (\lambda_1,\lambda_2;1) =
\sum_{\lambda_3,r_3} \tilde{K}_{r_1r_2 \lambda_3}^{
\lambda_1 \lambda_2 r_3} K_{\lambda_2 \lambda_1 r_3} ^{
r'_1 r'_2 \lambda_3} \tag semi$$ Equation \(semi) embodies
the fulfillment of the bootstrap program. This $R$--matrix
is the only one, among the one--parameter family of
semi-cyclic intertwiners \(rmat) satisfying the
Yang--Baxter equation, which does enjoy  the
``crossing''--symmetry   $$R_{r_1 r_2}^{r'_1 r'_2} (\xi_1 ,
\xi_2;1) = R_{r_2 r_1}^{r'_2 r'_1} (\xi_2 , \xi_1;1)
\eqno(13)$$ This noteworthy property is crucial for the
interpretation of $R$ as a scattering matrix.

  Compare, again, with the case of $q^p=1$
($p\ge3$): there, the unique intertwiner satisfying
Yang--Baxter always has crossing symmetry. The
crossing--summetric $R$--matrix \(semi) with the particular
choice  $\Omega(\xi)=1$ is thus the natural extension to
$p=2$ of the  general semi-cyclic solutions.  The  family of
$R$--matrices affinized by $\Omega(\xi)$ is peculiar to
$q=-1$, but we shall now study in detail the particular
crossing--symmetric $R$--matrix with $\Omega=1$.

\subhead{4. The particular solution: soliton interpretation}
\taghead{4.}
The non--zero
entries of the crossing--symmetric
$R(\xi_1,\xi_2;1)=R(\xi_1,\xi_2)$ are, explicitly,
$$\eqalign{ & R_{00}^{00}(\xi_1, \xi_2)=1 \cr  &
R_{01}^{01}(\xi_1, \xi_2)={\left(1-\lambda_1\right)
\left(1+\lambda_2\right) \over 1-\lambda_1\lambda_2} \cr  &
R_{01}^{10}(\xi_1, \xi_2)={\lambda_1-\lambda_2 \over
1-\lambda_1\lambda_2} \cr  & R_{10}^{01}(\xi_1, \xi_2)=
{\lambda_2-\lambda_1  \over 1-\lambda_1\lambda_2 } \cr &
R_{10}^{10}(\xi_1, \xi_2)= {\left(1+\lambda_1\right)
\left(1-\lambda_2\right)
 \over 1-\lambda_1\lambda_2}\cr
& R_{11}^{11}(\xi_1, \xi_2)=1 \cr}\eqno(rsem)$$

Due to the crossing symmetry \(13), in addition to the
already noted unitarity  and normalization properties
\(11-12), the ``semi-cyclic'' $R$--matrix admits a  clear
interpretation as a solitonic $S$--matrix: we may picture
the two states  $e_0(\lambda)$ and  $e_1(\lambda)$ as
localized around each one of the two potential minima, with
$\lambda$ a label very much like {\bf relativistic }
velocity.

Introduce the ``relative velocity'' $$u_{12}=
{\lambda_1-\lambda_2 \over 1-\lambda_1\lambda_2} $$ in
terms of which the ``semi-cyclic'' intertwiner reads as
$$R_{12}=\bordermatrix{  & 00 & 01 &10 & 11 \cr
  00 & 1 & 0 & 0 & 0 \cr
 01 & 0 & 1-u_{12} & u_{12} & 0 \cr
10 & 0 &-u_{12} & 1+ u_{12} & 0 \cr
 11& 0 & 0  & 0& 1 \cr}
\tag ddde$$ and the Yang--Baxter equation becomes
 $$R_{12}(u) R_{13}\left( {u+v \over 1+uv} \right)
R_{23}(v) = R_{23}(v) R_{13}\left( {u+v \over 1+uv}
\right)R_{12}(u) \tag$$  Note that only when $\Omega=1$ can
we parametrize the whole $R$--matrix in terms of a single
quantity $u_{12}$. In the usual trigonometric solutions to
the Yang--Baxter equations, the rapidities $u$ and $v$ add
up linearly, to $u+v$. Here the rapidities add up like
relativistic velocities!  It thus turns out that the labels
of the irreps under the quantum group may be identified
with  kinematical parameters:  the
two--dimensional Poincar\'e group is thus a manifestation
of an internal quantum symmetry. This situation at $p=2$
is very similar to the limit $p\to\infty$ of the general
semi-cyclic intertwiner \refto{12}.

 The braid group limit of the $R$--matrix \(ddde) is
obtained when $u\to\pm1$, \ie in the extreme relativistic
regime. Letting $R_\pm = \lim_{u\to\pm1} R(u)$, we find
$$R_+=\pmatrix { 1 & 0 & 0 & \cr 0 & 0 & -1 & 0 \cr 0 & 1
& 2 & 0 &\cr 0 & 0 & 0 & 1 \cr} \qquad
 R_-=\pmatrix { 1 & 0 & 0 & \cr 0 & 2 & 1 & 0 \cr 0 & -1
& 0 & 0 &\cr 0 & 0 & 0 & 1 \cr} \tag vfr$$ Let us
concentrate on one of them, say $R_-$ (the analysis is
identical for $R_+$).
It can be viewed as a particular case
of the more general $$ R_{(b,c)}=\pmatrix { 1 & 0 &
0 & \cr 0 & 1-bc & b & 0 \cr 0 & c & 0 & 0 &\cr 0 & 0 & 0 &
1 \cr} \tag $$
which satisfies Yang--Baxter without spectral
parameter and is thus a good starting point for the
construction of an extended Yang--Baxter system and hence a
link invariant \refto{13}. Indeed, we find that $\mu_0=bc$
and $\mu_1$=1 satisfy $$\eqalign{ &\left( \mu_i \mu_j -
\mu_k \mu_\ell \right) R_{ij}^{k\ell} =0 \cr & \sum_j
R_{ij}^{kj} \mu_j = AB \delta _i^k \cr
& \sum_j \left( R^{-1}\right)_{ij}^{kj} \mu_j =
A^{-1}B \delta _i^k \cr}\tag$$ with $A^{-1}=B= \sqrt{bc}$.
Accordingly, if $\alpha\in B_n$ is a word of the braid
group, the link invariant associated with its closure
$\hat\alpha$  is $$T(\hat\alpha) = A^{-w(\alpha)} B^{-n}
\tr \left( \rho(\alpha) \otimes \mu^{\otimes n} \right)
\tag$$ where $w(\alpha)$ is the wraith number of $\alpha$
and $\rho$ is the representation of the braid group
assigning to each generator $\sigma_i^{\pm1}$ the matrix
$R^{\pm1}_{(b,c)}$ acting on the $i$-th and $(i+1)$-th
strands. The link invariant satsifies the skein rule
$$A^2 P_+ - A^{-2} P_- = \left( A- A^{-1} \right) P_0
\tag$$ and is so normalized that $N$ disconnected unknots
are assigned the polynomial $(A+A^{-1})^N$. This is just
Jones polynomial in $t=A^2$. This is a nice result,
although the particular value $bc=-1$ in
\(vfr) is in fact singular: $T_{(1,-1)}(\hat\alpha)=0$ for
all $\alpha\in B_n$.

\subhead{5. Other cyclic solutions to Yang--Baxter}
\taghead{5.}

Have we found all the
$R$--matrices which  satisfy Yang--Baxter and intertwine
among cyclic representations of $U_q(s\ell(2))$ with
$q^2=-1$? This is interesting because it would represent a
major, if not final, step towards the classification of all
the $4\times4$ $R$--matrices satisfying Yang--Baxter. We
have not been able to prove that the family $R(\Omega)$
exhausts the solutions, although we strongly suspect that
this is indeed the case except for exceptional'' situations.
For example, a notable
 curiosity  of $U_q(s\ell(2))$ with $q=-1$ is that
there exists a parameter region $4\mu\nu=1$  for which the
raising and lowering generators coincide, up to a
proportionality factor: $F=2\nu E$. The quadratic Casimir
is then zero. Fixing furthermore $\nu=1$ (thus $\mu=1/4$)
allows us to find a different $R(\lambda_1, \lambda_2)$,
namely  $$R(\lambda_1, \lambda_2) = R(u_{12}) =
\pmatrix{ 1 & 0 & 0 & 0\cr    0 & u_{12} & 1-u_{12}& 0
\cr 0  & 0 & 1 & 0\cr 1-u_{12} &  0  & 0 &
 u_{12}\cr} \tag $$ with $$u_{12}= {\left( 1-
 \lambda_1 \right) \lambda_2 \over \lambda_1  \left( 1-
\lambda_2 \right) }\tag$$  This $R$--matrix (or its
transpose, another independent solution) is fundamentally
different from those found above in that the conservation
of quantum isospin holds only modulo 2:
$R_{00}^{11}\not=0$. Interpreting the $R$--matrix as
Boltzmann weights, we have  found thus a special case of the
eight--vertex model.

  We see also that the dependence of the ``relative
velocity'' $u_{12}$ on the quantum group labels $\lambda_1$
and $\lambda_2$ is now quite bizarre, although the
composition of spectral parameters in the Yang--Baxter
equation is  simply multiplicative: $u_{13} = u_{12}
u_{23}$.  The braid limit of this solution is singular.

\subhead{6. Conclusions and outlook}\taghead{6.}
 It is of course tantalizing to speculate on the dynamical
origin of space--time symmetries, albeit in the simplest
framework of 1+1 integrable systems. It is clear from our
analysis that the irrep label $\lambda$ can be understood
as the velocity of the state. Since the $R$--matrix
\(ddde) intertwines only irreps sharing the same values
of $\mu$ and $\nu$, these should be associated with some
extensive variables of the dynamical system. A clear
interpretation of them is likely to arise from the study
of the  (vertex-- or IRF--like) statistical--mechanical
model based on the $R$--matrices of this letter, which we
shall address elsewhere. A physical interpretation for the
affinization parameter $\Omega$ is also lacking.

The physics
of the $p=2$ case is the fermionic analogue of the bosonic
$p\to\infty$ limit considered in \refto{12}, and we expect
the intermediate cases $2<p<\infty$ to be associated with
anyonic statistics.
 We hope that the properties of the simple quantum group
studied in this letter will guide us towards a deeper
physical understanding of quantum groups at roots of unit
and, eventually, towards a classification of all
solutions to the Yang--Baxter equation.

 {\bf Acknowledgements}. I would like to thank C. G\'omez,
G.~Sierra and Ph.~Zaugg for interesting discussions.  This
work is partially supported by the Fonds National Suisse
pour la Recherche Scientifique.

\references

\refis{1} H. Au--Yang, B.M. McCoy, J.H.H. Perk, S. Tang and M.~Yan,
\pl,A123,219,1987&; B.M. McCoy, J.H.H. Perk, S. Tang and
C.H.~Sah, \pl,A125,9,1987&.

\refis{2}R.J. Baxter, J.H.H. Perk and H. Au--Yang,
\pl,A128,138,1988&; R. Baxter, \journal{J. Stat.
Phys.},57,1,1990&; R. Baxter, V.V. Bazhanov and J.H.H.
Perk, \ijmp,B4,803,1990&.

\refis{3}V.V.~Bazhanov and Yu.G.~Stroganov, \journal{J. Stat.
Phys.},51,799,1990&;  V.V.~Bazhanov, R.M.~Kashaev,
V.V.~Mangazeev and Yu.G.~Stroganov, \cmp,138,393,1991&;
V.V.~Bazhanov and R.M.~Kashaev, \cmp,136,607,1991&.

\refis{33}R.M. Kashaev, V.V. Mangazeev and Yu.G.
Stroganov, {\sl Cyclic eight--state $R$--matrix related to
$U_q(s\ell(3))$ at $q^2=-1$}, Protvino preprint (1991).

\refis{4} E. Date, M. Jimbo, M. Miki and T. Miwa,
\pl,A148,45,1990&; \cmp,137,133,1991&; RIMS preprints
703, 706, 715, 729 (1990).

\refis{5}C. de Concini and V. Kac,  \journal{Prog.
Math.},92,471,1990&; C.~de~Concini, V.~Kac and C.~Procesi,
Pisa preprint (1991) and RIMS preprint (1991).

\refis{6}C. G\'omez, M. Ruiz--Altaba and G. Sierra, \pl,B265,95,1991&..

\refis{88}V.). Tarasov, {\sl Cyclic monodromy matrices for
the $R$--matrix of the six--vertex model and the chiral
Potts model with fixed spin boundary conditions}, RIMS
preprint 774 (1991).

\refis{8}D. Bernard and V. Pasquier, {\sl Exchange algebra and exotic
supersymmetry in the chiral Potts model}, Saclay preprint SPhT/89--204.

\refis{9}D. Arnaudon, \pl,B268,217,1991&; CERN preprint TH.6324/91.

\refis{11}C. G\'omez and G. Sierra,  \np,B352,1991,791&;
C.~Ram\'{\i}rez, H.~Ruegg and M.~Ruiz--Altaba,
\np,B364,195,1991&.

\refis{12}C. G\'omez and G. Sierra, \pl,B,,1991&.

\refis{13}V.G. Turaev, \journal{Invent. Math.},92,527,1988&.

\refis{Ros}M. Rosso, Cuomo lectures (August 1991).

\endreferences
\vfill\eject
\endit

\endit